\documentclass[twocolumn,superscriptaddress]{revtex4}

\usepackage{graphicx}
\usepackage{xfrac}
\usepackage{verbatim}
\usepackage[unicode=true, bookmarks=true, bookmarksnumbered=false, bookmarksopen=false, breaklinks=false, pdfborder={0 0 1}, backref=false, colorlinks=true]{hyperref}

\usepackage[normalem]{ulem} %need for strikethrough
\usepackage{color} %needed for coloring of \todo etc.

\begin{document}

\title{Electromagnetically Induced Transparency in a mono-isotopic $^{167}$Er:$^7$LiYF$_4$ crystal below 1 Kelvin}

\author{N.~Kukharchyk}
\email[E-mail: ]{nadezhda.kukharchyk@physik.uni-saarland.de}
\affiliation{Experimentalphysik, Universit\"at des Saarlandes, D-66123 Saarbr\"{u}cken, Germany}

\author{D.~Sholokhov}
\affiliation{Experimentalphysik, Universit\"at des Saarlandes, D-66123 Saarbr\"{u}cken, Germany}

\author{O.~Morozov}
\affiliation{Kazan Federal University, 420008 Kazan, Russian Federation}

\author{S.~L.~Korableva}
\affiliation{Kazan Federal University, 420008 Kazan, Russian Federation}

\author{A.~A.~Kalachev}
\affiliation{RFC Kazan Scientific Center of RAS, 420029 Kazan, Russian Federation}

\author{P.~A.~Bushev}
\affiliation{JARA-Institute for Quantum Information (PGI-11), Forschungszentrum Jülich, 52428 Jülich, Germany}

\date{\today}

\begin{abstract}
%Reliable quantum memory is a necessary component for realization of long-distance quantum communication protocols and architecture.

Electromagnetically induced transparency allows for controllable change of absorption properties which can be exploited in a number of applications including optical quantum memory.
In this paper, we present a study of the electromagnetically induced transparency in $^{167}$Er:$^6$LiYF$_4$ crystal at low magnetic fields and ultra-low temperatures.
Experimental measurement scheme employs optical vector network analysis which provides high precision measurement of amplitude, phase and pulse delay. 
%paves the way towards full on-chip integration of optical quantum memory setups.
%Experimentally derived values of the spin width and pulse delay are analysed with respect to predictions of available theoretical model.
We found that sub-Kelvin temperatures are the necessary requirement for studying electromagnetically induced transparency in this crystal at low fields. 
A good agreement between theory and experiment is achieved taking into account the phonon bottleneck effect.

\end{abstract}
\maketitle

%\pacs{42.50.Fx, 76.30.Kg, 03.67.Hk, 03.67.Lx, 76.30.-v}
%\keywords{Cooperative phenomena in quantum optical systems, Superconducting qubits, Quantum communication, Quantum computation architectures and implementations, EPR in condensed matter}
\section{Introduction}
Electromagnetically induced transparency (EIT) is a quantum interference effect which can be observed in a multilevel atomic system where the interference of two different excitation pathways leads to appearance of a narrow transparency window inside the atomic absorption spectrum \cite{Harris1997,Fleischhauer2005}. Such a controllable change of absorption properties, which is accompanied by a strong dispersion and enhanced nonlinearity, can be used in a number of fascinating applications including slow light \cite{Kash1999,Hau1999,Budker1999}, optical storage \cite{Liu2001,Phillips2001} and quantum memory (QM) \cite{Chaneliere2005,Eisaman2005}. In particular, the latter is currently considered as a basic ingredient for long-distance quantum communication via quantum repeaters and for scalable optical quantum computers, which makes it a topic of active research \cite{Bussieres2013,Heshami2016}. In this respect, significant experimental progress has been achieved in demonstrating EIT-based QM in cold atomic ensembles \cite{Chen2013,VernazGris2018,Hsiao2018,Wang2019}. On the other hand, among the most discussed systems suitable for memory implementation are solids doped with rare-earth (RE) ions \cite{Thiel2011}, which exhibit extremely long spin coherence times at low temperatures \cite{Zhong2015}, high optical densities, and compatibility with photonic integrated circuits \cite{Zhong2017}.  Moreover, solids doped with Kramers RE ions (ions having even number of electrons) provide optical transitions inside telecom bands thereby making frequency conversion unnecessary, and hyperfine transitions in 1-10 GHz range that is compatible with microwave quantum photonics. Therefore, studying EIT in these materials is of interest for both developing optical QM and interfacing to superconducting quantum devices.

%Regarding solids doped with RE ions, 
EIT has been extensively studied in the crystals doped with praseodymium ions \cite{Ham1997,Ichimura1998,Turukhin2002,Longdell2005,Akhmedzhanov2006,Goldner2009,Heinze2013,Schraft2016,Fan2019}, but much less in those doped with Kramers RE ions such as erbium and neodymium \cite{Baldit2010,Akhmedzhanov2018}. In the present work, we report the first observation of EIT in a mono-isotopic $^{167}$Er$^{3+}$:$^7$LiYF$_4$ (Er:LYF) crystal. Isotopically purified LYF crystals, where only the $^7$Li isotope is present, have long been known for their ultra-narrow optical inhomogeneous broadening ($\sim 10$~MHz) \cite{Macfarlane1992,Macfarlane1998,Thiel2011,Kukharchyk2018}. Due to this feature, the hyperfine structure of the optical transitions can be clearly resolved so that observing EIT needs no special ensemble preparation using spectral hole burning. For the same reason, the system is very attractive for implementing off-resonant Raman quantum memory protocols \cite{Moiseev2011,Moiseev2013,Zhang2013,Kalachev2013,Zhang2014}, and studying EIT in this crystal may be considered as the first step towards their realisation.

The optical/spin coherence properties of LYF system were studied in 1990’s \cite{Ganem1991,Meltzer1992}. Today, LYF crystals are again in the focus of spectroscopic research, but now within the context of QM applications \cite{Gerasimov2016, Marino2016,Akhmedzhanov2016,Kukharchyk2017,Akhmedzhanov2018}. It turns out that Er:LYF system is a quite challenging material, because it requires high magnetic field $B\gtrsim1.5$~T at low temperature $T\sim1.5$~K for establishing optical coherence time at $\sim10~\mu$s timescale~\cite{Ganem1991,Meltzer1992,Marino2016}.  In our recent work, however, we have demonstrated an increase of optical coherence time by deep freezing of the crystal to ultra-low temperatures, i.e., $T\ll 1$~K~\cite{Kukharchyk2017}, and by taking advantage of an optical clock transition that appears at a low magnetic field. In the present work, we use the same approach with respect to spin transitions of the impurity ions, which makes it possible to observe EIT in Er:LYF crystal below 1 Kelvin by identifying a $\Lambda$-system with a zero-first-order-Zeeman (ZEFOZ) spin transition. In doing so, we use the optical vector network analysis (OVNA) technique~\cite{Capmany2013,Kukharchyk2018}, which provides high-precision measurements of absorption and dispersion properties of the EIT medium.

\begin{figure*}[ht!]
\includegraphics[width=1\textwidth]{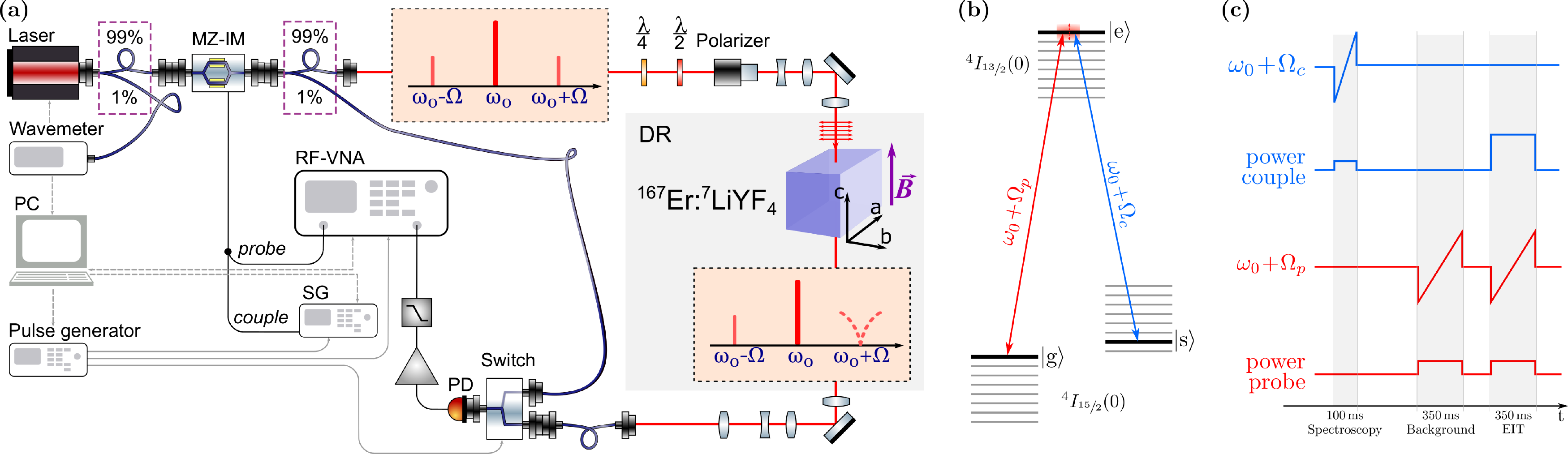}
\caption{(Color online) (a) Illustration of the experimental setup. MZ-IM stands for the Mach-Zehnder intensity modulator. DR is the dilution refrigerator. PD is the high-speed InGaAs photoreceiver. RF-VNA is the radio-frequency vector network analyser. SG is the RF signal generator. Both RF-VNA and SG are triggered by using a pulse generator. PC stands for the controlling computer. %The experiment is controlled by using a PC.
(b)~Schematics of the energy-level structure indicating levels invoved in the EIT measument. (c)~Measurement sequence for EIT consists of three steps: spectroscopy on the pump transition, background and EIT measurements.}
\label{Setup}
\end{figure*}

\section{Experimental setup}

We investigate a single Er:LYF crystal doped with 0.0025\% atomic concentration of $^{167}$Er$^{3+}$ ions. The crystal is grown by Bridgman-Stockbarger method as described in \cite{Kukharchyk2017} and has dimensions of 5~mm x 5~mm x 5~mm. The crystal axis $c$ is directed along the applied magnetic field $B$ and along the light propagation direction, see Fig.~\ref{Setup}(a). The crystal is thermally anchored to the mixing chamber of the dilution refrigerator by using a silver-based conducting glue, see Ref.~\cite{Kukharchyk2017} for further detail.

For the optical spectroscopy and for the observation of EIT we employ OVNA setup, as outlined in Fig.~\ref{Setup}(a).
%The schematics of the experimental setup is outlined in Fig.~\ref{Setup}(a).
The erbium doped free running fiber laser (NKT Photonics Adjustik E15) emits a continuous signal at $\omega_0/2\pi = 195888~$GHz, which corresponds to the transition $^4I_{15/2}(0)-{}^4I_{13/2}(0)$ of the $^{167}$Er impurity ions. The laser field is polarized perpendicular ($\sigma$) to the crystal $c$ axis. The laser frequency is stabilized by using an optical wavelength-meter (High-Finesse WS6-200).
Signals emitted by a radio-frequency vector network analyzer (RF-VNA) and a radio-frequency signal generator (SG) modulate the laser field via a Mach-Zehnder intensity modulator (MZ-IM). This scheme employs the homo-heterodyne measurement technique, as described in \cite{Kukharchyk2018}. The modulated optical signal contains three frequencies  (carrier and two sidebands) out of which only one sideband interacts with the crystal. The carrier frequency is far-detuned from the absorption feature. The resulting beat signal is then detected by a high speed InGaAs photoreceiver, filtered on undesired higher order beats, amplified and returned to the VNA. The background is analysed by putting a small portion of the modulated light directly to the photoreceiver.

The described OVNA method allows to simplify construction of the setup for heterodyne detection and guarantees a phase match of the sidebands with the main frequency, where the last one serves as the heterodyne. The VNA provides the fast and direct extraction of the amplitude, phase and electrical delay of the signal with a high precision. Implementation of this method narrows down the gap between the optical and microwave fields. Though in general this method requires a post-processing of the measured signal, see \cite{Kukharchyk2018}, in case of small amplitudes and phases, the signal can be treated as undistorted with a simple polynomial conversion of the absorption dB-signal into optical depth.

The measurement sequence for the EIT contains three data-readout steps, see Fig.~\ref{Setup}(c).
First step is spectroscopy on the $|s\rangle\leftrightarrow|e\rangle$ coupling transition, see Fig.~\ref{Setup}(b) for energy level schematics, which allows to obtain the frequency of the coupling transition via fitting the spectral line and thus allows to minimize the detuning $\Delta_{se}$ of the coupling transition at each measurement step. Following after several seconds delay on the device reconfiguration, the absorption on the probe transition $|g\rangle \leftrightarrow |e\rangle$ is first measured with the couple laser off and with the probing intensity of $\sim$0.2~W\,cm$^{-2}$, which is the second step of the measurement sequence. At the third step, the second absorption profile on $|g\rangle \leftrightarrow |e\rangle$ is measured with the coupling beam of $\sim$6.5~W\,cm$^{-2}$ intensity on the $|s\rangle \leftrightarrow |e\rangle$ transition. %\textit{With that we can estimate Rabi frequencies for the probe transition, $\omega_{\textrm{ge}}\simeq2\pi\cdot0.9$~MHz, and for the coupling transition, $\omega_{\textrm{se}}\simeq2\pi\cdot5.4$~MHz.}

\section{Optical spectroscopy}

\begin{figure*}[t!]
	\includegraphics[width=2\columnwidth]{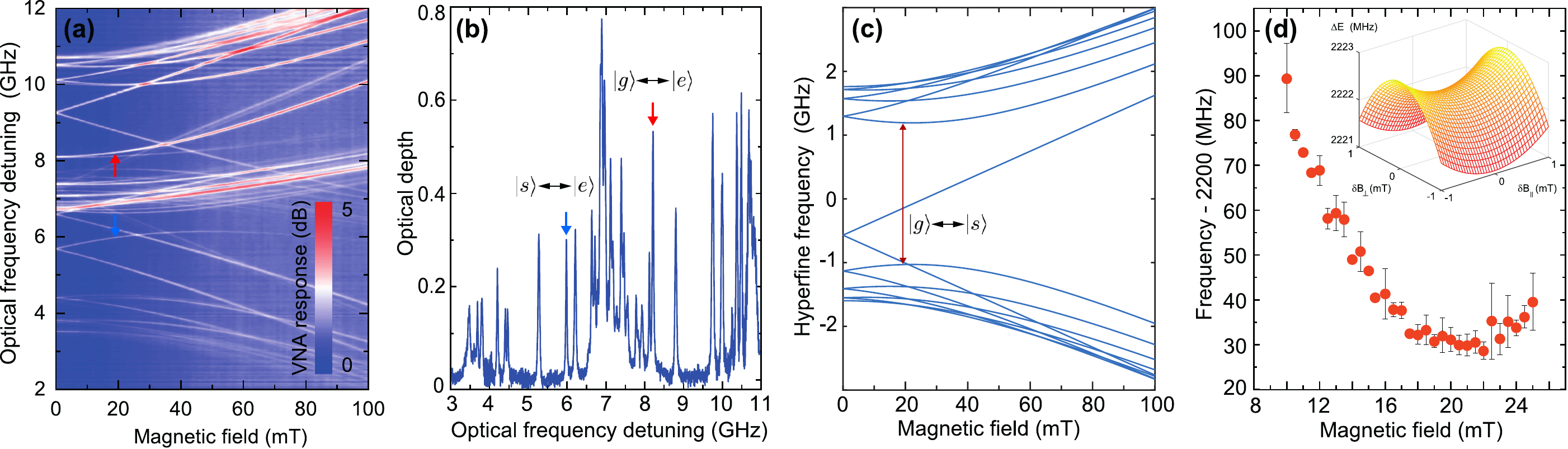}
	\caption{(Color online) (a)~Absorption spectrum of the sample on the ${}^4I_{15/2}(0)-{}^4I_{13/2}(0)$ transition as a function of the longitudinal magnetic field and~(b)~absorption spectrum at the magnetic field of 20~mT. The frequency is given in detuning from the 196.888~THz of the main laser frequency. The absorption lines $|g\rangle\leftrightarrow|e\rangle$ and $|s\rangle\leftrightarrow|e\rangle$ form a symmetrical $\Lambda$-structure used for EIT.
		(c)~Calculated hyperfine structure of the electronic ground state $^4I_{15/2}(0)$ as a function of the longitudinal magnetic field. The red arrow indicates the ZEFOZ point of spin transition between the states $|g\rangle$ and $|s\rangle$, which is observed at the magnetic field of 20~mT.
		(d)~Measured transition frequency between the hyperfine states $|g\rangle$ and $|s\rangle$ as a function of the longitudinal magnetic field around the ZEFOZ point. The inset shows the modelled hyperfine transition frequency dependence on the longitudinal and transverse magnetic field detunings from the ZEFOZ point.}
	\label{Spectra}
\end{figure*}

The absorption spectrum of $^4I_{15/2}\leftrightarrow{}^4I_{13/2}$ transition is measured as a function of the longitudinal magnetic field in the range from 0 to 100 mT at the base temperature of the dilution refrigerator, shown in Fig.~\ref{Spectra}(a). Three groups of lines can be identified at higher fileds. Two groups at frequencies below 5~GHz and above 9~GHz correspond to the two transitions with spin-flip,
$^4I_{15/2}(0)\vert-1/2\rangle \leftrightarrow {}^4I_{13/2}(0)\vert +1/2 \rangle $
and
$^4I_{15/2}(0)\vert +1/2 \rangle \leftrightarrow {}^4I_{13/2}(0)\vert -1/2 \rangle $
respectively.
The third, central, group of lines covering frequencies between 5~GHz and 9~GHz has small g-factors and corresponds to the transitions without spin-flip, $^4I_{15/2}(0)\vert\pm1/2\rangle \leftrightarrow {}^4I_{13/2}(0)\vert \pm1/2 \rangle $.

The structure of the optical spectrum is determined by the Zeeman splitting of the ground- and excited-state Kramers doublets of $^{167}$Er and hyperfine interaction. To simulate it, we take advantage of the effective spin Hamiltonian for the ground and excited states:
\begin{eqnarray}\label{SpinHam}
H & = & g_{\parallel}\mu_B B_z S_z +g_{\perp}\mu_B(B_xS_x+B_yS_y)\nonumber\\ & &+AI_zS_z+B(I_xS_x+I_yS_y)\nonumber\\& &+P[I_z^2-I(I+1)/3],
\end{eqnarray}
where $\mu_B$ is the Bohr magneton; $g_\parallel$  and $g_\perp$  are the components of the $g$ factor parallel and perpendicular to the $c$ axis, respectively; $B_i$ are the components of the external magnetic field; $S_i$ and $I_i$ are the electron- and nuclear-spin operators, respectively; $A$ and $B$ are the hyperfine parameters; and $P$ is the quadrupole constant. Here $S=1/2$ is the effective electron spin, and $I=7/2$ is the nuclear spin for $^{167}$Er. The ground state parameters have long been known from EPR measurements \cite{Sattler1971}: $A=-325.8$~MHz, $B=840$~MHz, $g_{\perp}=8.105$ and $g_{\parallel}=3.137$. The excited state parameters can be determined from the measured optical spectra by a fitting procedure as was done in \cite{Macfarlane1992,Macfarlane1998}. As a result, for the excited state $^4I_{13/2}$ we obtain $A=-170$~MHz, $B=970$~MHz, $P=15$~MHz, and $g_{\parallel}=1.56$. The value of $g_{\parallel}$ is close to the previously estimated values of 1.52~\cite{Gerasimov2016} and 1.6~\cite{Marino2016}.

In possession of the effective spin Hamiltonian parameters for the ground and excited states, we are able to find an appropriate $\Lambda$ structures of optical transitions for observing EIT. For the longest spin coherence times to be achieved, it is desirable to use the zero first order Zeeman (ZEFOZ) transitions \cite{Fraval2004,Lovric2011,McAuslan2012,Zhong2015,Ortu2017}, while the most efficient Raman interaction requires symmetrical $\Lambda$ structure. Numerics show that both conditions can be satisfied in the present crystal at the longitudinal magnetic field of about 20~mT. In this case, a couple of hyperfine sublevels of the ground state, namely
\begin{eqnarray}
\vert s \rangle  & = & \frac{1}{\sqrt{2}} \left( \vert -1/2, 7/2\rangle + \vert 1/2, 5/2 \rangle \right), \\
\vert g \rangle  & = & \frac{1}{\sqrt{2}} \left( \vert -1/2, 7/2\rangle - \vert 1/2, 5/2 \rangle \right)
\label{States1}
\end{eqnarray}
(here $\vert m_s, m_I \rangle$ are the eigenstates of the Hamiltonian in Eq.~\ref{SpinHam}), form ZEFOZ transition, while the hyperfine sublevel of the excited state
\begin{eqnarray}
\vert e \rangle  & = & \vert 1/2, 7/2\rangle
\label{States2}
\end{eqnarray}
is equally coupled to only these states thereby providing a symmetric and isolated $\Lambda$-structure.

Absorption spectrum at 20~mT is demonstrated in Fig.~\ref{Spectra}(b), where optical transitions forming the $\Lambda$-structure are marked by red arrows.
The observed difference of optical depth is due to equilibrium population distribution between sublevels $|g\rangle$ and $|s\rangle$ at low temperature. The linewidth of both optical transitions is about 32~MHz.

The simulated structure of the spin-levels is plotted in Fig.~\ref{Spectra}(c), indicating the point of ZEFOZ transition by red arrow. 
Frequency of the microwave transition  $|g\rangle \leftrightarrow |s\rangle$ is extracted from the frequency difference of the probe and couple transitions at each magnetic field value is plotted in Fig.~\ref{Spectra}(d). We observe the expected ZEFOZ point at approximately 20~mT, where the microwave frequency takes minimal value. 
The inset shows the simulated spin transition frequency in dependence on the detuning of the magnetic field from the ZEFOZ point, where longitudinal, $\delta B _{\parallel}$, and transverse, $\delta B _{\perp}$, detunings are along and perpendicular to the $c$ axis, respectively.

\section{Electromagnetically induced transparency}

\begin{figure}[t!]
	\includegraphics[width=1\columnwidth]{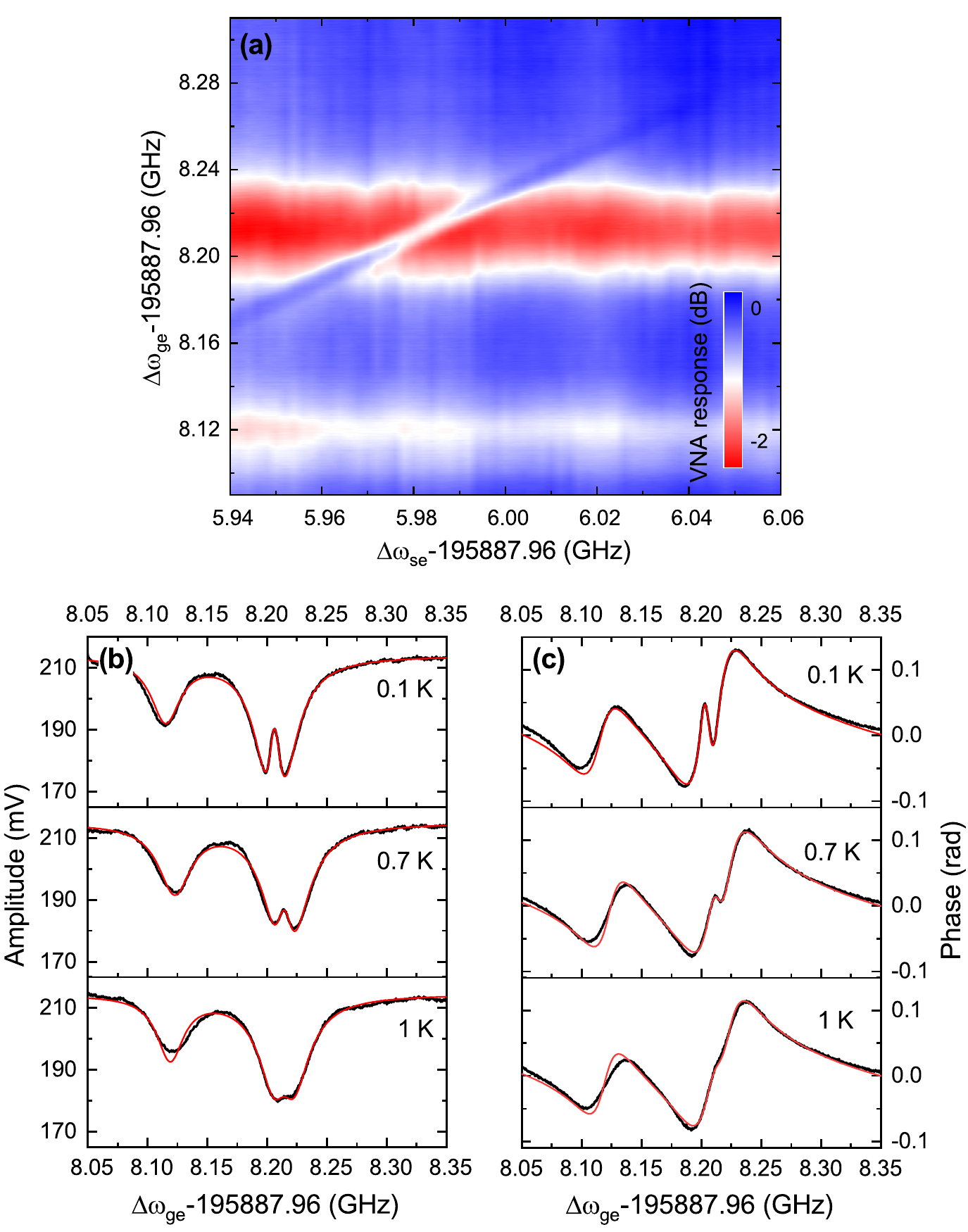}
	%Fig3_1
	\caption{(Color online) (a) Plot of EIT as a function of the couple frequency at 20~mT. The probe and couple frequencies are given in detunings from the 195.888 THz of the main laser frequency.
		Slices of the measured amplitude (b) and phase (c) of the EIT signal are shown for three temperatures: 0.1~K, 0.7~K and 1~K. The black line corresponds to the experimental signal, the red line shows the fit of the data to the OVNA model~\cite{Kukharchyk2018}, Eqs.~(\ref{eq_OVNA1}-\ref{eq_OVNA}), with the spectral shape given by Eq.~(\ref{eit_func}).}
	\label{EIT_spectra}
\end{figure}

The EIT feature appears when the coupling frequency is swept through the resonance, as shown in Fig.~\ref{EIT_spectra}(a). The probe transition appears as wide red line at approx.~8.21~GHz. At 8.12~GHz we have another week absorption line which is close to the probe transition and is, thus, fit together with it. The observed EIT resonance, as shown in Fig.~\ref{EIT_spectra}(b,c), vanishes with increase of temperature and does not reveal any superhyperfine structure.
Previously, the superhyperfine structure was found in the Electron Paramagnetic Resonance spectra of Nd:LiYF$_4$ crystals~\cite{Aminov2013}.
Recently, splitting of the EIT line into 9 lines due to the superhyperfine interaction has been observed by Akhmedzhanov et. al.~\cite{Akhmedzhanov2018} in monoisotopic Nd:LiYF$_4$. While Nd and some other rare earth ions, e.g. Ce and Yb, revealed superhyperfine structure in LiYF$_4$ and other substrates, observation of superhyperfine structure in Er:LiYF$_4$ spectra has not been reported yet.
We did not observe any signature of the superhyperfine interaction in the EIT feature as well. This is explained by large broadening of the spin line, $\simeq$5~MHz,  which cannot be further reduced even when working at the clock transition and sub-Kelvin temperatures and, thus, leads to the smearing out of the superhyperfine structure.

We derive the amplitude and the phase of the EIT spectrum by fitting the experimental data to the OVNA model as explained in \cite{Kukharchyk2018}. Absorption depth $\alpha L(\omega)$ and phase $\phi(\omega)$ of the EIT spectral feature are derived as imaginary, $\alpha L=\Im[\chi]$, and real, $\phi=\Re[\chi]$, parts of the susceptibility. We then fit the experimental data to the OVNA model~\cite{Kukharchyk2018} as follows:
\begin{eqnarray}
\alpha L & \propto & \left( 1 + \Im[\chi]^2 + 2\Im[\chi]\cos(\Re[\chi])^2 \right)^{\sfrac{1}{2}},
\label{eq_OVNA1}
\\
\phi & \propto & - { \Im[\chi]\sin(\Re[\chi]) \over 1 + \Im[\chi]\cos(\Re[\chi]) }.
\label{eq_OVNA}
\end{eqnarray}

The imaginary and real parts of the susceptibility are derived from \cite{Goldner2009}:
\begin{eqnarray}
\chi = i { \lambda \alpha_0 \over 2 \pi } &
{ \Gamma_{\textrm{ge}}(\Gamma_{\textrm{sg}}+i(\Delta\omega_{\textrm{ge}}-\Delta\omega_{\textrm{se}})) \over
	(\Gamma_{\textrm{ge}} + i\Delta\omega_{\textrm{ge}})(\Gamma_{\textrm{sg}}+i(\Delta\omega_{\textrm{ge}}-\Delta\omega_{\textrm{se}}))+|{\Omega_{\textrm{c}} \over 2 }|^2}
\label{eit_func},
\end{eqnarray}
where $\Delta\omega_{\textrm{ge}}$ and $\Delta\omega_{\textrm{se}}$ are the detunings of the probe and coupling beams, respectively; $\Omega_{\textrm{c}}$ is the coupling Rabi frequency. $\Gamma_{\textrm{ge}}$ and $\Gamma_{\textrm{sg}}$ are the linewidths of the excited, $|e\rangle$, and auxiliary, $|s\rangle$, states, respectively, and comprise of the natural linewidth $\sfrac{1}{\textrm{T}_1}$, dephasing rate $\sfrac{1}{\pi \textrm{T}_2}$, and the inhomogeneous broadening $\Gamma_{\textrm{inh}}^{\textrm{ge(sg)}}$ as follows:
\begin{eqnarray}
\Gamma_{\textrm{ge(sg)}}& = {1 \over \textrm{T}_1^{\textrm{ge(sg)}}} + {1 \over \pi \textrm{T}_2^{\textrm{ge(sg)}}} + \Gamma_{\textrm{inh}}^{\textrm{ge(sg)}} = \sfrac{1}{2}~\Gamma_{\textrm{opt}(\textrm{HF})}.
\end{eqnarray}
We further use the notation $\Gamma_{\textrm{opt}}$ and $\Gamma_{\textrm{HF}}$, which are the full width at half maximum (FWHM) of the optical $|g\rangle \leftrightarrow |e\rangle$ and spin $|s\rangle \leftrightarrow |e\rangle$ transitions, respectively.
The dephasing component of the both decay rates, $\sfrac{1}{\pi \textrm{T}_2^{\textrm{ge(sg)}}}$, is then sensitive to the direct and indirect flip-flops and to the direct and phonon bottleneck processes.

%\textcolor{blue}{\textbf{Open question (important for the discussion of the results):} Which contribution to the FWHM is the dominating one? Assuming that the inhomogeneous linewidth is independent of temperature and field, we end up with $\gamma_{\textrm{ge(sg)}}=\sfrac{1}{T_1}$ and $\gamma_{\textrm{ge(sg)}\,deph}=\sfrac{1}{\pi T_2^*}$ being sensitive to the flip-flop and bottleneck.}

We have measured the EIT feature in the range of magnetic field from 10~mT to 25~mT and in the temperature range from 100~mK to 1~K at the magnetic field of 19~mT.
Fitting the experimental data to Eqs.~(\ref{eq_OVNA1}-\ref{eit_func}) allows to extract the optical $\Gamma_{\textrm{opt}}$ and spin $\Gamma_{\textrm{HF}}$ FWHM linewidth, and coupling strength $\Omega_{\textrm{c}}$.
The group delay $\tau_d$ is derived directly by the RF-VNA as $\partial\phi / \partial \omega$. Examples of the fits of the absorption amplitude and phase profiles at three selected temperatures are shown in Fig.\ref{EIT_spectra}(b,c).
Fitting of the transition at $\sim$8.12~GHz simultaneously with the EIT-line allows for better precision of the derived parameters.

\begin{figure}[t!]
	\includegraphics[width=1\columnwidth]{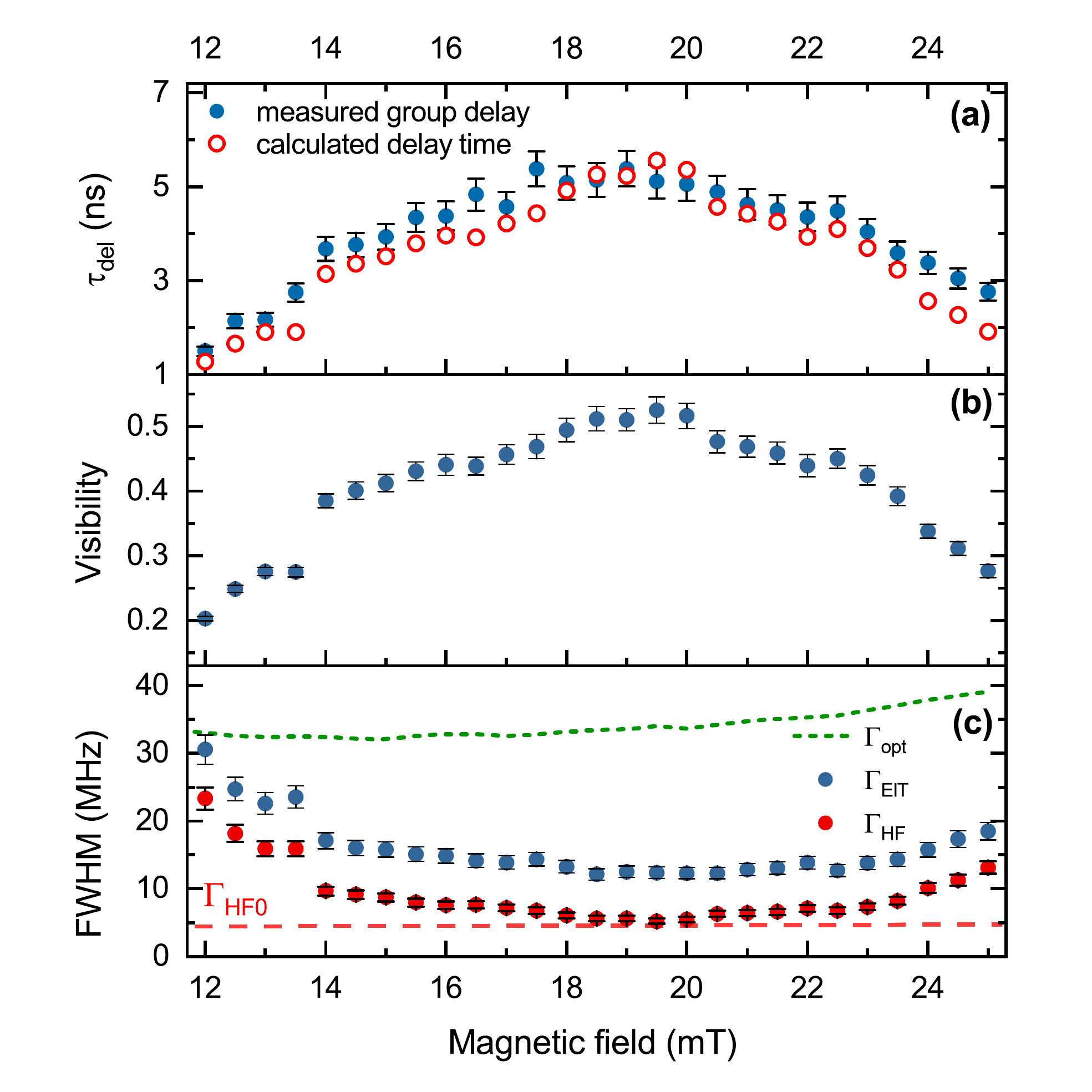}
	%Fig3_1
	\caption{(Color online) (a)~VNA-measured (blue dots) and calculated (red circle) group delay as a function of the magnetic field. (b)~Dependence of the EIT visibility on the magnetic field, derived with Eq.~(\ref{Eq_EIT_Vis}). (c)~Optical, $\Gamma_{\textrm{opt}}$, and microwave, $\Gamma_{\textrm{HF}}$, linewidths as derived from the experimental data. Width of the EIT transparency window, $\Gamma_{\textrm{EIT}}$, is calculated with Eq.~(\ref{Eq_EIT_Width}). $\Gamma_{\textrm{HF0}}$ is the guiding line for the minimal width of the spin transition at the ZEFOZ point.}
	\label{EIT_versus_B}
\end{figure}

Extracted optical and spin linewidths are shown in Fig.~\ref{EIT_versus_B}(c) and Fig.~\ref{EIT_versus_T}(c). The optical linewidth $\Gamma_{\textrm{opt}}$ does not depend on temperature and shows weak broadening with an increase of the magnetic field: at 19~mT it equals $\Gamma_{\textrm{opt}}\simeq2\pi\times 35$~MHz and increases up to $\Gamma_{\textrm{opt}}\simeq 2\pi\times40~$MHz at 25~mT.
Such increase is related to the inhomogeneity of the g-factor, $\Delta g(\omega)$, which leads to an associated broadening $\Delta g(\omega)\!\cdot\!B$.
On change of the magnetic field, $\Gamma_{\textrm{HF}}$ takes minimal value in the vicinity of the ZEFOZ point. Converse to the $\Gamma_{\textrm{HF}}$, the group delay $\tau_d$ and visibility increse when approaching ZEFOZ point.
The coupling frequency $\Omega_{\textrm{c}}$ is independent of the magnetic field and temperature and equals $\Omega_{\textrm{c}}\simeq 2\pi\times15~$MHz, which proves that we work in the pure EIT regime, $\Gamma_{\textrm{opt}}>\Omega_{\textrm{c}}$.

%The visibility $\textrm{V}_{\textrm{EIT}}$, spin linewidth $\Gamma_{\textrm{HF}}$, EIT transparency width $\Gamma_{\textrm{EIT}}$ and group delay $\tau_D$, also have an extremum point at about 19~mT, see Fig.~\ref{EIT_versus_B}.

%\section{Magnetic field dependence of the spin width}
%\begin{eqnarray}
%\label{Eq_noise_gen}
%&\Gamma_{\textrm{HF}}(\Delta B)&= \Gamma_{spin0} + \\
%&&\sum_{i=x,y,z} S_{1i}\delta B_i + S_{2i} \delta B_i \sqrt{2\delta B_i^2+4\Delta B^2}\nonumber,
%\end{eqnarray}

%Width of the spin transition $\Gamma_{\textrm{HF}}$ takes the minimal value of $\simeq2\pi\cdot5.2$~MHz at $\sim19$~mT.

%the dephasing rate and, consequently, the spin width are governed by amplitudes  and rates of magnetic field fluctuations rising from electron spins of the dopant ions and nuclear spins of the host material (predominantly fluorine)~\cite{Kukharchyk2017}. These fluctuations are typically described by the flip-flop rates of nuclear and electronic spins which are known to have $\sim sech^2(\sfrac{\sfrac{\gamma}{2\pi}\, B}{k_BT})$ dependence~\cite{Abragam1970} on the magnetic field $B$ and temperature $T$, where $\sfrac{\gamma}{2\pi}$ is the gyromagnetic ratio of the spin transition in \sfrac{Hz}{T}, and $k_B$ is the Boltzmann constant.

In the vicinity of the ZEFOZ transition, the dependence of $\Gamma_{\textrm{HF}}$ on decoherence and relaxation processes is overshadowed by the sensitivity of the spin system to the fluctuations of the magnetic field amplitude caused by the flip-flops, which is given by gradient, $S_{1i}=\partial \omega / \partial B_i$, and curvature, $S_{2i}=\partial^2 \omega / \partial^2 B_i$, of the spin transition frequency $\omega$ in the magnetic field $B$~\cite{Kukharchyk2017}:
\begin{equation}
\Gamma_{\textrm{HF}}(\Delta B)= \Gamma_{\textrm{HF0}} +S_{1}\delta B + S_{2} \delta B \sqrt{2\delta B^2+4\Delta B^2},
\label{eq:spin_width_dB}
\end{equation}
where $\Delta B$ is the detuning of the magnetic field from the ZEFOZ point, and $\Gamma_{\textrm{HF0}}$ is the detuning-independent inhomogeneous broadening of $|s\rangle \leftrightarrow |g\rangle$ microwave transition. The amplitude of the magnetic noise, $\delta B$, is the total amplitude of the external magnetic field fluctuations experienced by the erbium ions over the time of the measurement.
%In our experiment, each measurement point is averaged over a period of $\simeq 200~\mu$s, thus the sensed magnetic noise contains full amplitudes for frequencies $f_{noise} \gtrsim 5$~kHz and partial amplitudes for frequencies $f_{noise} \lesssim 5$~kHz. %All slower fluctuations are only partially included in $\delta B$.
Equation~(\ref{eq:spin_width_dB}) does not include any linear inhomogeneous spin broadening associated with the g-factor inhomogeneity, $\Delta g(\omega)\!\cdot\!B$~\cite{Kukharchyk2017}. We can nevertheless assume that it is similar to the broadening of the optical linewidth $\Gamma_{\textrm{opt}}$ and becomes relevant only at the magnetic fields above 23~mT, see Fig.~\ref{EIT_versus_B}(c).
Values of the curvature, $S_2 = 0.91$~\sfrac{MHz}{mT$^2$}, and gradient, $S_1 = \Delta B \cdot 0.91$~\sfrac{MHz}{mT$^2$}, are extracted from the absorption spectra of the spin transition, see Fig.~\ref{Spectra}(d). Inserting these values into Eq.~(\ref{eq:spin_width_dB}), we derive the minimal broadening of the spin transition, $\Gamma_{\textrm{HF0}} \simeq 2\pi \times4.5$~MHz, and the magnitude of the magnetic noise, $\delta B_{\textrm{ZEFOZ}} \simeq 0.4$~mT.

%Similar to the magnetic noise values, the spin linewidth $\Gamma_{\textrm{HF}}$ strongly decreases between 10~mT and 14~mT, see Fig.~\ref{EIT_versus_B}c.
%In Kukharchyk et. al.\cite{Kukharchyk2017}, we have estimated the amplitudes of the magnetic field fluctuations induced by the electron and nuclear spins, which is $\simeq 10~\mu$T in the 0.005$\%$-doped sample and $\simeq 3~\mu$T in the 0.0013$\%$-doped sample at the magnetic field $\sim 40$~mT and for the averaging period of $\lesssim 100~\mu$s. We see that at twice lower magnetic field and at least twice larger spectral width, the magnetic noise amplitude increases by approx. two orders of magnitude. Such a difference can be related to much weaker spin polarization, higher effective temperature and reduction of the frozen core effect in current experiment at 19~mT. Rapid drop of the magnetic noise amplitude at (10-12)~mT can also signify the beginning of formation of the frozen core.

In order to obtain the visibility of the EIT signal, $\textrm{V}_{\textrm{EIT}}$, and the width of the EIT transparency window, $\Gamma_{\textrm{EIT}}$, we use the theory published in Refs.~\cite{Kuznetsova2002,Fan2019}, in which the standard model in Eq.~(\ref{eit_func}) is generalized for the inhomogeneously broadened solid-state system. 
%Major assumption taken by Kuznetsova et. al. implies $\Gamma_{\textrm{HF}}<<\Gamma_{\textrm{opt}}$, which holds for the most of the magnetic field and temperature ranges in our experiment.
For large inhomogeneous spin broadening, the linewidth of the EIT feature is larger than the spin linewidth and depends on the ratio of the square of coupling strength to the product of optical and spin broadenings~\cite{Kuznetsova2002}:
\begin{eqnarray}
%\Gamma_{\textrm{EIT}} & =& \Gamma_{\textrm{HF}} \sqrt{ 2t(1+t)\left(1+\sqrt{1+ {(1+\sfrac{1}{2}t^{-1})^2 \over (1+t)^2}}\right)   },\nonumber \\
%t&=&{ \Omega_{\textrm{c}}^2 \over 2{opt}\Gamma_{\textrm{HF}} }
\Gamma_{\textrm{EIT}}&= &\Gamma_{\textrm{HF}}\left(1+{ \Omega_{\textrm{c}}^2 \over \Gamma_{\textrm{opt}}\Gamma_{\textrm{HF}} }\right)
\label{Eq_EIT_Width}
\end{eqnarray}
The EIT linewidth is thus broader than the spin width by the add-up of $\Omega_{\textrm{c}}^2 / \Gamma_{\textrm{opt}} \simeq 2\pi \times 6.7$~MHz.
%\textcolor{blue}{\textbf{Remark:} Im most papers inhomogeneous broadening is the doppler broadening. So I find it important to point out, that this model is meant for the solid state.}
The EIT visibility is derived as \cite{Kuznetsova2002}:
\begin{eqnarray}
\textrm{V}_{\textrm{EIT}} & = &{\Omega_{\textrm{c}}^2 \over \Omega_{\textrm{c}}^2 + \Gamma_{\textrm{opt}}\Gamma_{\textrm{HF}}}.
\label{Eq_EIT_Vis}
\end{eqnarray}
In the vicinity of the ZEFOZ point, $\simeq19$~mT, we have $\Omega_{\textrm{c}}^2 \sim \Gamma_{\textrm{opt}}\Gamma_{\textrm{HF}}\simeq (2\pi\times16$~MHz$)^2$ which gives maximum visibility of $\simeq0.55$. We observe the visibility reduction when detuning the magnetic field from the ZEFOZ point, Fig.~\ref{EIT_versus_B}(b), and with the increase of the temperature, Fig.~\ref{EIT_versus_T}(b), so that the EIT feature is well-visible below 0.7~K, see Fig.~\ref{EIT_spectra}(b).
In order to attain $\textrm{V}_{\textrm{EIT}}\rightarrow 1$, the coupling strength must be much larger than the optical and microwave inhomogeneities, $\Omega_{\textrm{c}}^2 \gg \Gamma_{\textrm{opt}}\Gamma_{\textrm{HF}}$, which is not feasible at our experimental condition.

\section{Temperature dependence of the spin width}

\begin{figure}[ht!]
\includegraphics[width=1\columnwidth]{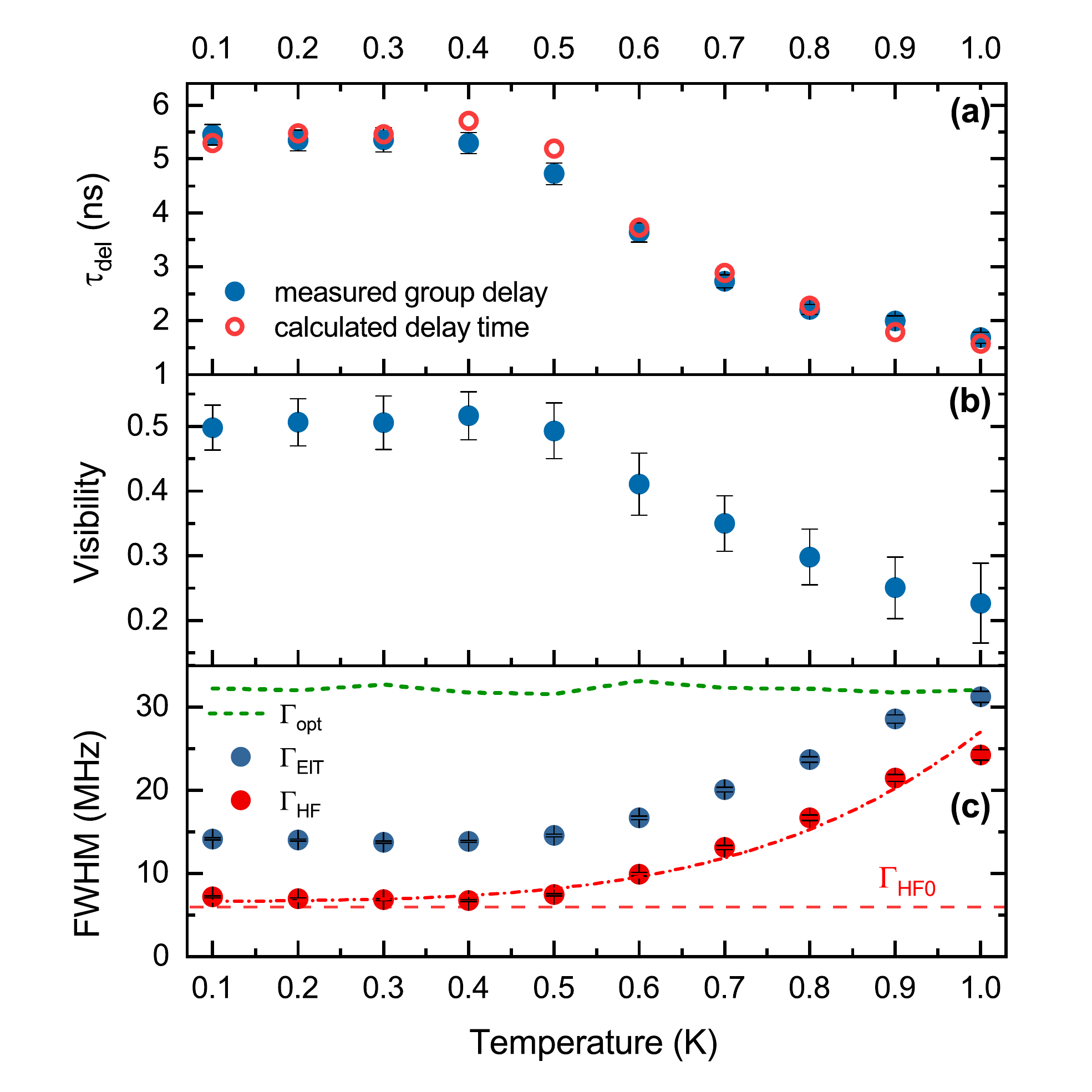}
\caption{(Color online) (a)~VNA-measured (blue dots) and calculated (red circle) group delay as a function of the temperature.
	(b)~Dependence of the EIT visibility on the temperature, derived with Eq.~(\ref{Eq_EIT_Vis}).
	(c)~Width of optical and spin transitions and EIT transparency window as a function of temperature. Fit of $\Gamma_{\textrm{HF}}$ to the broadening from non-equilibrium phonons, Eq.~(\ref{eq_T_bn_fit}), is shown with dash-dotted red line. From the NQP-model, $\Gamma_{\textrm{HF0}}$ is the minimal attainable width of spin transition at the ZEFOZ point.
}
\label{EIT_versus_T}
\end{figure}

Temperature dependence of the EIT parameters is measured in the range from 0.1~K to 1~K at 19~mT and is plotted in Fig.~\ref{EIT_versus_T}. Width of the optical transition remains constant in the full range of temperatures, $\Gamma_{\textrm{opt}}\simeq 2\pi \times 32$~MHz, see Fig.~\ref{EIT_versus_T}(c).
Width of the spin transition increases with the increase of temperature, while visibility decreases, as shown in Fig.~\ref{EIT_versus_T}(b).

The spin linewidth $\Gamma_{\textrm{HF}}$ decreases with the decrease of temperature till it saturates below 0.5~K, which is associated with reaching the minimal attainable temperature of the sample, $\textrm{T}_{\textrm{min}}$. 
The temperature dependence of $\Gamma_{\textrm{HF}}$ below 1~K can be governed by spin-flip process, direct process and phonon bottleneck process. 
In some works, the phonon bottleneck (PB) has already been observed when studying optical coherence~\cite{Kukharchyk2019} and spin relaxation~\cite{Budoyo2018} phenomena in Er:Y$_2$SiO$_5$ below 1~K. 
%In particular, in the work by Budoyo et. al.~\cite{Budoyo2018}, strong phonon bottleneck was a dominating process in the temperature dependence of the relaxation time at a spin transition. In our study of optical coherence in $^{167}$Er:Y$_2$SiO$_5$, contribution from PB was very strong though not the dominating one, and it was accompanied by the flip-flop process~\cite{Kukharchyk2019}. 
In the previous study of optical coherence in $^{166}$Er:LiYF$_4$ below 1~K~\cite{Kukharchyk2017}, in the absence of a reliable model for effective temperature, it was not possible to extract the flip-flop and bottleneck rates from the experimental data.

According to the present study, the dependence of the $\Gamma_{\textrm{HF}}$ on temperature is dominated by the bottleneck process, which is associated with the presence of non-equilibrium phonons (NQP). Both NQP and PB describe influence of the same physical phenomenon on relaxation and coherence in the spin ensemble. Speaking of the line-broadening due to the PB, we stick to the term of the non-equilibrium phonons, which are the actual source of the effects.
Line broadening due to the NQP can be described as~\cite{Graf1998,Abragam1970}:
\begin{eqnarray}
\Gamma_{\textrm{NQP}}&=%&{\sigma \upsilon \over \pi} \Sigma_{ph}\nonumber\\
 {\sigma \upsilon \over \pi} {2 \omega^2 \Delta\omega \over 2 \pi \upsilon^3} \coth \left({\hbar \omega \over \textrm{k}_\textrm{B}\textrm{T}}\right)^2,
\label{eq_T_bn}
\end{eqnarray}
where $\sigma\sim100$~nm$^2$ is the collision cross-section of the phonons~\cite{Wu2013,Graf1998}, $\upsilon\simeq 5.5$~\sfrac{km}{s}~\cite{Minisini2005}, $\omega$ is the frequency of the microwave transition, $\Delta\omega$ is the linewidth of the spin transition, and $\sfrac{\hbar\omega}{\textrm{k}_\textrm{B}}\simeq0.05$~K. Based on Eq.~(\ref{eq_T_bn}), we fit $\Gamma_{\textrm{HF}}$ to the following expression:
\begin{eqnarray}
\Gamma_{\textrm{HF}}&=& \Gamma_{\textrm{HF0}} +\Gamma_{\textrm{HF0}} \,\gamma_{\textrm{NQP}}\coth \left({\hbar \omega \over \textrm{k}_\textrm{B} \textrm{T}_{\textrm{eff}}}\right)^2,
\label{eq_T_bn_fit}
\end{eqnarray}
where $\Gamma_{\textrm{HF0}}$ is the NQP-independent linewidth, $\gamma_{\textrm{NQP}}=\sfrac{\sigma \omega^2 }{ \pi^2 \upsilon^2}\sim 10^{-3} - 10^{-5}$ is the dimensionless coefficient.

In our previous studies of optical coherence in Erbium-doped crystals below 1~K, we found that the temperature in vicinity of the excited Erbium ensemble differs from what we measure on the temperature sensor on the cryostat. Therefore, we introduce an effective temperature, $\textrm{T}_{\textrm{eff}}$, which is the true temperature of the spin-ensemble and is given by~\cite{Kukharchyk2019}
\begin{equation}
\textrm{T}_{\textrm{eff}}=\textrm{T}_{\textrm{min}}(1+\sfrac{\textrm{T}}{\textrm{T}_{\textrm{min}}})^{\sfrac{1}{2}},
\label{eq_tmin}
\end{equation}
where $\textrm{T}_{\textrm{min}}$ is the minimal temperature attained by the sample at particular experimental conditions.

The fitting results are shown in Fig.~\ref{EIT_versus_T}(c). The obtained parameters agree quite well with the expected values: $\Gamma_{\textrm{HF0}}\simeq 2\pi\times6.4$~MHz, $\gamma_{\textrm{NQP}}\simeq2\cdot10^{-3}$, and $\textrm{T}_{\textrm{min}}\simeq0.5$~K. 
%For the reliable analysis of the data and extraction of precise parameters we would require analysis of the magnetic field dependence on flip-flop rate and NQP contribution which is not feasible in the EIT experiment at the ZEFOZ point.
%We nevertheless can expect that considering the flip-flop will introduce a minor correction to the temperature dependence of $\Gamma_{\textrm{HF}}$.
Current results suggest that presence of NQP is major limiting factor for coherence at milli-Kelvin temperatures.

\section{Group delay}
Dispersive properties of an EIT medium are attrctive in application to the variable time-delays and the on-demand storage and retrieval of optical pulses~\cite{Novikova2012,Fleischhauer2005}. These is possible due to reduction of the group velocity of light when EIT is achieved.
In OVNA experiment, the group delay $\tau_d$ is directly calculated by the VNA: for each frequency segment, VNA detects change of the phase of the received signal with respect to the initial phase. This method offers a straightforward way of direct identification of the delay associated with the slowdown of the light pulses in EIT medium.

\begin{comment}
	The group delay is known to be larger for smaller coupling frequency $\Omega_{\textrm{c}}$ and reads as :
	\begin{equation}
	\tau_d = {L \over \upsilon_g} = \alpha L {\Gamma_{\textrm{opt}} \over 2 \Omega_{\textrm{c}}^2},
	\end{equation}
	where $\alpha L$ is the absorption depth of the optical transition that can be found from the absorption spectrum ($\alpha L \simeq 0.5$, see Fig~\ref{EIT_spectra}(b)); $L$ is the length of the crystal; $\upsilon_g$ is the group velocity. We found that the maximum value of the group delay is $\tau_d \simeq 5.7~\textrm{ns}$, which correlates with the measured delay at the ZEFOZ point.
\end{comment}

Slow light is associated with the concept of dark polaritons: simultaneous propagation of the light and spin waves with reduced speed~\cite{Fleischhauer2002}. The contribution of spin component in the polariton state is proportional to the collective light-ion coupling strength $g\sqrt{\textrm{N}}$, which is estimated as $g\sqrt{\textrm{N}} = \mu\sqrt{\sfrac{\omega \textrm{n}_{\textrm{at}}}{2 \hbar \epsilon_0}}\simeq2\pi\times270$~MHz, where the transition dipole moment equals $\mu~\simeq~2.5~\cdot~10^{-32}$~Cm, $\omega$ is the optical frequency of the probe field, and $\textrm{n}_{\textrm{at}}~\simeq~7~\times~10^{15}$~cm$^{-3}$ is the concentration of the atoms.
The relative contribution of the optical component is described by so called mixing angle, which reads as $\theta \simeq \arctan{\sfrac{g^2\textrm{N}}{\Omega_{\textrm{c}}^2}}$ and equals to $\simeq 85^{\circ}$ under present experimental conditions, which means that most of the energy is concentrated in the spin component~\cite{Fleischhauer2002}.
The delay time estimated in terms of polariton propagation is $\tau_\textrm{d} = {\textrm{L} \over c} {g^2\textrm{N} \over \Gamma_{\textrm{opt}}\Gamma_{\textrm{HF}}}$ \cite{Fleischhauer2002,Kuznetsova2002}, where $c$ is the speed of light in vacuum. Calculated $\tau_\textrm{d}$ values are presented in Fig.~\ref{EIT_versus_B}(a) and Fig.~\ref{EIT_versus_T}(a). We find very good correlation of the experimental and theoretical values.
%We however must note that tolerance of the $g\sqrt{N}$ is of $\pm2\pi\times50~$MHz which leads to that of the calculated delay time of $\pm3~$ns, which still leaves us in a good agreement with the experimental result.

\section{Conclusion}

In conclusion, we have presented a detailed study of electromagnetically induced transparency (EIT) in mono-isotopic $^{167}$Er:LiYF$_4$ crystal measured at sub-Kelvin temperatures.
%The $\Lambda$-scheme for EIT employs ZEFOZ on the spin transition.
We have found that the EIT process at low magnetic fields can only be observed at milli-Kelvin temperatures.
The detected transparency window has a minimal FWHM linewidth of 12~MHz and maximal transparency of 50\% at the minimal attainable temperature of 500~mK. Measured experimental parameters agree with the theoretical model involving the phonon bottleneck effect.

Optical vector network analysis approach is successfully employed in the measurement scheme and is thus proved as an efficient method of measuring the electromagnetically induced transparency. It can be further deployed in other quantum memory schemes.

Though cooling down below 1~K allowed to suppress most of the decoherence processes, we found that efficiency of the EIT is limited by rather high minimal attainable temperature, $\simeq$500~mK. Such elevated effective temperature is due to high optical excitation power as well as to poor thermal interface and thermal conductance of the crystal medium below 1~K. As the consequence, dependence of the spin width on the temperature is limited by dynamics of non-equilibrium phonons.

Upon improvement of the thermal interface between sample and cryostat, it is possible to have a further enhancement of the coherence properties while working at low magnetic field. Also, Nd:LiYF crystal would be another interesting candidate for low-temperature experiments due to observation of the hyperfine structure. This will allow to develop reliable quantum memory feasible for low-magnetic-field applications.

\section{Acknowledgement}

This work is supported by DFG through the grant BU~2510/2-1. A.K. acknowledges financial support from the Government of the Russian Federation (Mega-Grant No. 14.W03.31.0028). P.B. acknowledges Jürgen Eschner and Giovanna Morigi for the valuable discussions.

%\section{Appendix I: Fitting model}
%Intensity of the VNA signal is fit to~\cite{Kukharchyk2018}:

\bibliography{EIT1}

\end{document}